\documentstyle[preprint,psfig,aps]{revtex}

\def\be{\begin{equation}}
\def\ee{\end{equation}} 
\def\ba{\begin{eqnarray}}
\def\ea{\end{eqnarray}} 
\newcommand{\ignore}[1]{} 
\begin{document}
\draft
\title{Invariant Manifolds and Collective Coordinates}
\author{T. Papenbrock$^a$ and T. H. Seligman$^b$}
\address{Centro Internacional de Ciencias, Cuernavaca, Morelos, 
Mexico\\
and\\
$^a$ Institute for Nuclear Theory,
University of Washington, Seattle, WA 98195 USA\\
$^b$ Centro de Ciencias F\'\i sicas, 
University of Mexico (UNAM), Cuernavaca, Mexico}
\maketitle
\begin{abstract}
We introduce suitable coordinate systems for interacting many-body systems with
invariant manifolds. These are Cartesian in coordinate and momentum space and
chosen such that several components are identically zero for motion on the
invariant manifold. In this sense these coordinates are collective.  We make a
connection to Zickendraht's collective coordinates and
present certain configurations of few-body systems where rotations and
vibrations decouple from single-particle motion.  These configurations do not
depend on details of the interaction.
\end{abstract}

\section{Introduction}
Dynamical systems with invariant manifolds in phase space have been the subject
of ongoing research in recent years. Many authors have considered the case of
two or more coupled identical systems that are chaotic. On invariant
manifolds the subsystems display identical or synchronized motion, and the
manifold's transverse stability is investigated
\cite{Caroll94,Brown94,Rulkov95,Beigie95,Gauthier96,Ott,Yang}.

An alternative approach is based on the observation that any rotationally
invariant system of identical interacting particles possesses low dimensional
invariant manifolds in classical phase space \cite{PS96}. On such manifolds,
the classical motion displays largely collective behavior and decouples from
more complex single-particle behavior. The importance of a given invariant
manifold depends crucially on its stability properties.  If the manifold under
consideration is sufficiently stable in transverse directions, the quantum
system may exhibit wave function scarring \cite{Wintgen,TP97,PP99} or display a
strong revival for wave-packets localized to the vicinity of the manifold
\cite{PSW98}.  These findings may be directly associated with the slow decay of
collective motion due to the coupling between collective and single-particle
motion.

In this paper we propose suitably adapted coordinate systems that separate
collective and single-particle motion on the invariant manifolds mentioned
above.  Such coordinates clarify the separation of collective and single
particle motion and may be useful in several applications. We have in mind (i)
the problem of damping and dissipation of collective excitations and the
interplay of collective and chaotic motion in atomic nuclei 
\cite{Guhr89,Blocki92,Heiss94,Wambach94,Bauer94,Manfredi95,Zelevinski96,TP00},
which is often addressed in the framework of single-particle motion in a 
time dependent mean field; 
(ii) multi particle fragmentation of atoms at threshold which evolves over
highly symmetric configurations corresponding to invariant manifolds
\cite{Wintgen,PattardRost98}; (iii) the structural stability of invariant
manifolds \cite{BJPS98}.

There is a traditional way to introduce collective and single-particle
coordinates in interacting many-body systems. Aiming at the description of
nuclear vibrations and rotations, Zickendraht \cite{Zickendraht} introduced a
system of collective coordinates in a self-bound many-body system. Three of
these coordinates describe the center of mass motion, and six collective
degrees of freedom govern the dynamics of the inertia ellipsoid. The remaining
coordinates are of single-particle nature. We shall establish the relation of
coordinates of invariant manifolds to those defined by Zickendraht. Furthermore
we shall show that more complicated collective motion, e.g. shearing modes can
be described

This article is divided as follows. In the next section we introduce suitable
coordinate systems for interacting many-body systems with invariant manifolds.
We give a construction recipe and present a detailed example calculation. As an
application we give a potential expansion around an invariant manifold and
discuss stability properties.  In section~\ref{zick} we make a connection with
the Zickendraht coordinates. We present examples where
the motion of the inertia ellipsoid corresponds to the motion on an invariant
manifold. For such initial conditions the traditional collective motion
decouples completely from the single-particle degrees of freedom.  We also find
that collective coordinates as defined here are capable of other types of
motion. Therefore we finally discuss how motion on or near such invariant
manifolds could be interpreted as collective motion of a system.

\section{Coordinates for invariant manifolds}
In this section we present a transformation from Cartesian single particle
coordinates in position and momentum space to Cartesian coordinates that are
adapted to invariant manifolds. The new coordinates consist of ``collective
coordinates'' that govern the motion on the invariant manifold and of
coordinates transversal to this manifold that represent the single-particle
aspects.

Consider rotationally invariant systems of $N$ identical particles in $d$
spatial dimensions ($d=2$ or $d=3$).  The Hamiltonian is invariant under both,
the action of the rotation group O($d$) and the group of permutations
S$_N$. One may now take a finite subgroup ${\cal G} \subset $ O($d$) with
elements $g$ and properly chosen permutations $P(g)$ such that 
\be
\label{mandef}
g P^{-1}(g) (\vec{p},\vec{q}) = (\vec{p},\vec{q}), \quad \forall g \in {\cal G}
\qquad  
\vec{p}\equiv (p_1,\ldots,p_{Nd}), \vec{q}\equiv (q_1,\ldots,q_{Nd})
\ee
for points $(\vec{p},\vec{q})$ on some invariant submanifold of phase space.
On such a manifold, the action of certain rotations $g$ can be canceled by
permutations. These permutations clearly form a subgroup isomorphic to 
${\cal G}$.

Fig.~\ref{fig1} shows a configuration of four particles in two spatial
dimensions that corresponds to a point on an invariant manifold. The operations
of elements from the discrete symmetry group ${\cal G}=C_{2v}$ can be undone by
suitable permutations of particles.  This leads to a collective motion with two
degrees of freedom which we shall identify with vibrations.

Fig.~\ref{fig2} shows two spatial configurations of eight (\ref{fig2}a) and six
(\ref{fig2}b) particles, respectively which display a $D_{4h}$ symmetry.  If
initial momenta display the same symmetry the motion on the invariant manifold
will have two degrees of freedom. For eight particles the radii of the two
circles will oscillate synchronously, and the two circles will vibrate against
each other. For the six particles we will have a vibration of the radius of the
circle and of the two particles along the vertical axis. We may choose initial
momenta to reduce the symmetry group to $C_{4h}$ which will allow rotations
around the vertical axis and thus add an additional degree of freedom.  For
eight particles we could alternatively choose initial conditions that are
limited to a $D_4$ symmetry. Besides the vibrations discussed above this would
allow for a shearing motion of the two circles thus yielding again three
degrees of freedom. We could also reduce the fourfold rotation axis to a
twofold one and obtain $D_{2h}$, $D_2$ or $C_{2v}$ as remaining symmetry groups
yielding more collective degrees of freedom. Adding two particles symmetrically
onto the principal axis of rotation would also increase the number of degrees
of freedom by one. Other reductions of symmetry will yield different invariant
manifolds with varying degrees of freedom. We will see this exemplified by
explicit construction of coordinates.

We may use the definition (\ref{mandef}) directly for the construction of
coordinate systems where invariant manifolds correspond to coordinate axis or
planes, i.e. non collective coordinates vanish for motion on the invariant
manifold. To this purpose we consider the many-body system in Cartesian
coordinates in momentum and position space. In what follows we will introduce
orthogonal transformations in configuration space only; momenta will be subject
to the same transformation. 

In a Cartesian coordinate system each element $g\in {\cal G}$ and each
permutation $P(g)$ can be represented by an orthogonal matrix ${\bf M}_g$ and
${\bf P}_g$ of dimension $Nd$.  It is clear that the products ${\bf M}_g{\bf
P}^{\rm T}_g$ form a matrix group ${\cal H}$ that acts onto position and
momentum space, respectively.  The construction of the coordinate system is now
straightforward. Every vector $\vec{p}$ and $\vec{q}$ may be expanded in basis
vectors of the irreducible representations (IRs) of ${\cal H}$ by means of
projectors \cite{Hammermesh}.
\be
\label{projec}
\Pi_\nu=\sum_{g\in{\cal G}} \chi_g^{(\nu)}{\bf M}_g{\bf P}^{\rm T}_g.
\ee
Here $\chi_g^{(\nu)}$ denotes the character of $g$ in the $\nu$'th IR.
Similar formulae hold for momentum space. The projection onto the identical IR
defines the invariant manifold. Note that the identical representation is
one-dimensional while the invariant manifolds of interest typically have higher
dimensionality. We can find independent vectors on the manifold by projecting
from different vectors, but in practice the construction of the independent
vectors seems to be unproblematic as we shall see in the example.

A comment on the rotation symmetry is in order. Like any Cartesian coordinates,
the coordinates introduced in this article do not explicitly reflect the
invariance under rotations.  Acting on an invariant manifold, rotations
generate a continuous family of equivalent manifolds. Our coordinates, however,
single out one particular manifold. For quantum systems, the rotation operator
may easily be constructed and used for projection onto subspaces of definite
angular momentum.

\section{A simple example}

\begin{figure}
  \begin{center}
    \leavevmode
    \parbox{0.9\textwidth}
    {\psfig{file=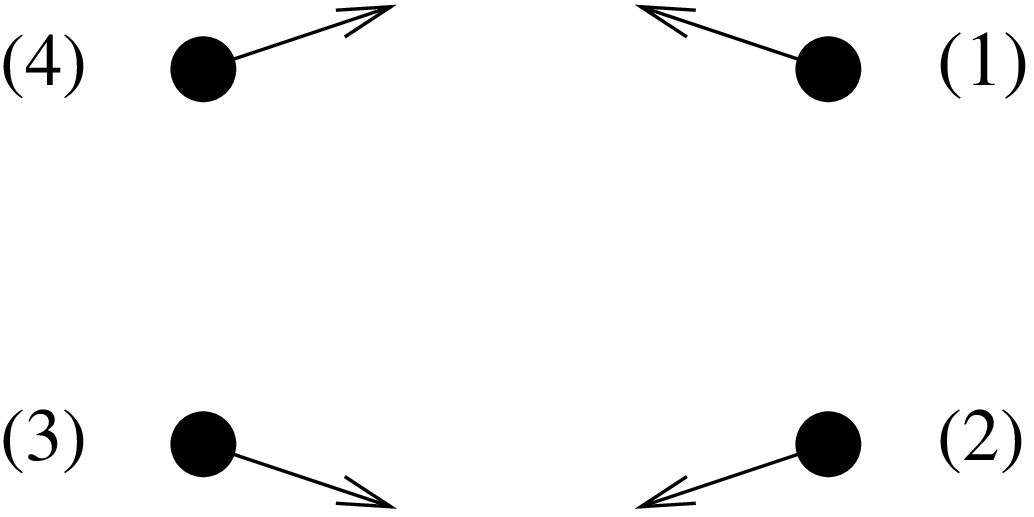,width=0.9\textwidth,angle=270}}
  \end{center}
  \protect\caption{Collective configuration on invariant manifold. Positions 
are indicated by filled circles and momenta by arrows.}
  \label{fig1}
\end{figure}

\begin{figure}
  \begin{center}
    \leavevmode
    \parbox{0.9\textwidth}
    {\psfig{file=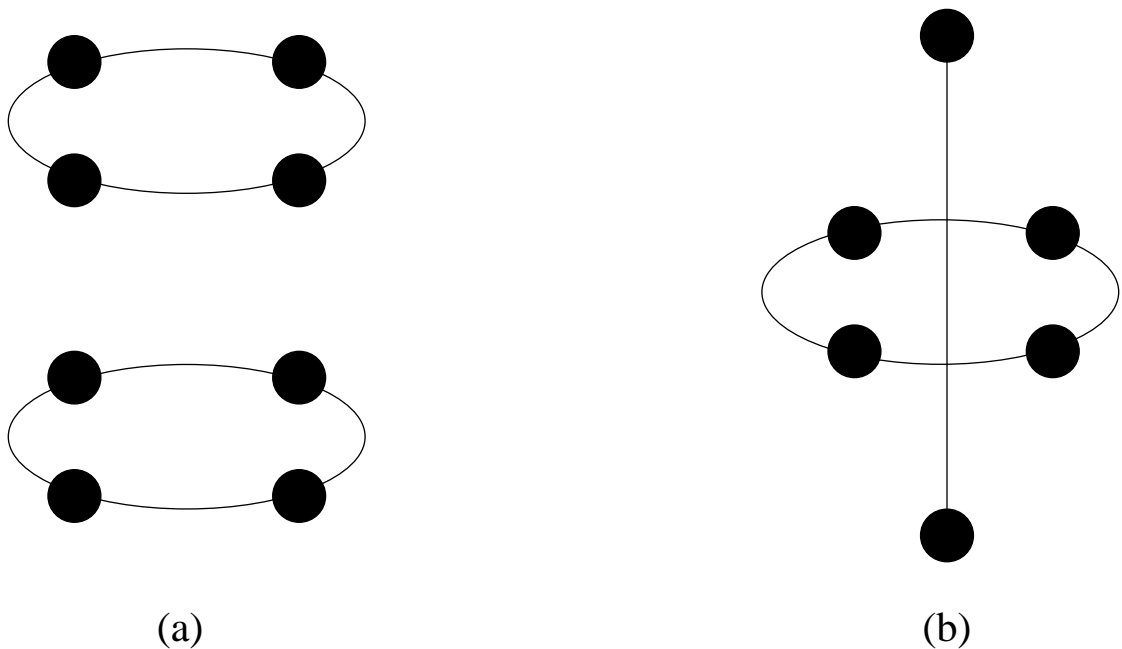,width=0.9\textwidth,angle=0}}
  \end{center}
  \protect\caption{Configurations of eight (a) or six (b) particles in 
    three dimensions that correspond to invariant manifolds. Positions are 
    indicated by filled circles.}
  \label{fig2}
\end{figure}

We now illustrate the proposed construction explicitly for four particles in
two dimensions and a quartic potential, considering the invariant manifold
shown in Fig.~\ref{fig1}. We shall also expand the potential near the invariant
manifold to second order in the transversal coordinates.

The invariant manifold is defined by those points which are invariant under
${\cal H}=\{E,\sigma_x P_{(12)(34)},\sigma_y P_{(14)(23)},C_2 P_{(13)(24)}\}$,
where $E$ denotes the identity, $P$ a permutation of particles as indicated,
$\sigma$ a reflection at the axis indicated, and $C_2$ a rotation about
$\pi$. Thus, ${\cal H}=C_{2v}$ with four IRs labeled by $\nu=A_1,B_1,A_2,B_2$
\cite{Hammermesh}. Let $\vec{q}=(x_1,x_2,x_3,x_4,y_1,y_2,y_3,y_4)$ denote a
coordinate vector in position space ($x_i,y_i$ denote the coordinates of the
$i$'th particle). We have
\ba 
E\,\vec{q} &=& (x_1,x_2,x_3,x_4,y_1,y_2,y_3,y_4),\nonumber\\ \sigma_x
P_{(12)(34)}\,\vec{q} &=& (x_2,x_1,x_4,x_3,-y_2,-y_1,-y_4,-y_3) \nonumber\\ 
C_2P_{(13)(24)}\,\vec{q} &=& (-x_3,-x_4,-x_1,-x_2,-y_3,-y_4,-y_1,-y_2) 
\nonumber\\
\sigma_y P_{(14)(23)}\,\vec{q} &=& (-x_4,-x_3,-x_2,-x_1,y_4,y_3,y_2,y_1)
\nonumber.  
\ea
Using the character table of $C_{2v}$ \cite{Hammermesh} and the projectors
(\ref{projec}) one constructs the following basis vectors corresponding to the IR labeled by 
\ba 
A_1&:& \qquad
e'_1=(1,1,-1,-1,0,0,0,0)/2, \quad e'_2=(0,0,0,0,1,-1,-1,1)/2, \nonumber\\
B_1&:& \qquad e'_3=(1,1,1,1,0,0,0,0)/2, \quad e'_4=(0,0,0,0,1,-1,1,-1)/2,
\nonumber\\ A_2&:& \qquad e'_5=(1,-1,-1,1,0,0,0,0)/2, \quad
e'_6=(0,0,0,0,1,1,-1,-1)/2, \nonumber\\ B_2&:& \qquad
e'_7=(1,-1,1,-1,0,0,0,0)/2, \quad e'_8=(0,0,0,0,1,1,1,1)/2.\nonumber 
\ea 
The vectors associated with the identical IR $A_1$ span the two-dimensional
invariant manifold and the vectors associated with the IRs $B_1,A_2,B_2$ span
the transverse directions.

We now present the orthogonal transformation that transforms the single
particle coordinates $\vec{q}$ into the coordinates adapted to the invariant
manifold. In our example $x$ and $y$-components do not mix and we have
\ba
\label{trans}
\left[\begin{array}{c} x_1' \\ x_2' \\ x_3' \\ x_4' \end{array}\right] = 
{1\over 2} \left[\begin{array}{rrrr} 1 &  1 & -1 & -1 \\
                                     1 &  1 &  1 &  1 \\
                                     1 & -1 & -1 &  1 \\
                                     1 & -1 &  1 & -1 \end{array}\right]
\left[\begin{array}{c} x_1 \\ x_2 \\ x_3 \\ x_4 \end{array}\right],
\qquad   
\left[\begin{array}{c} y_1' \\ y_2' \\ y_3' \\ y_4' \end{array}\right] = 
{1\over 2} \left[\begin{array}{rrrr} 1 & -1 & -1 &  1 \\
                                     1 & -1 &  1 & -1 \\
                                     1 &  1 & -1 & -1 \\
                                     1 &  1 &  1 &  1 \end{array}\right]
\left[\begin{array}{c} y_1 \\ y_2 \\ y_3 \\ y_4 \end{array}\right].
\ea

To illustrate the example and to further demonstrate the usefulness of the
newly introduced coordinate system we want to consider the the interacting
four-body system with Hamiltonian
\be
\label{ham}
H=\sum_{i=1}^4\left((p_{x_i}^2+p_{y_i}^2)/2 + 16(x_i^2+y_i^2)^2\right)
-\sum_{i<j}\left[(x_i-x_j)^2+(y_i-y_j)^2\right]. 
\ee
This Hamiltonian has been studied previously \cite{PSW98}. In particular,
the stability of the invariant manifold displayed in Fig.~\ref{fig1} has
been studied by computing the full phase space monodromy matrix of several
periodic orbits that are inside the invariant manifold. It was found that 
several orbits are linearly stable in transverse directions or possess rather
small stability exponents. Qualitatively, this may also be understood by
studying the Hamiltonian (\ref{ham}) close to the invariant manifold. 
We therefore use the transformation (\ref{trans}) and expand the potential
of Hamiltonian (\ref{ham}) to second order in the transverse directions
labeled by ($\epsilon_1,\ldots\epsilon_6$) while keeping the full dependence 
of the coordinates ($x,y$) inside the invariant
manifold. One obtains the quadratic form 
$\vec{\epsilon}^{\rm T} {\bf V} \vec{\epsilon}$
where
\ba
{\bf V}= \left[\begin{array}{cccccc} 12x^2+4y^2 & 0 & 0 &  16xy & 0 & 0 \\
                          0 & 12x^2 & 0 & 0 & 8xy & 0 \\
                          0 & 0 & 24x^2+8y^2 & 0 & 0 & 16xy \\
                          16xy & 0 & 0 & 8x^2+24y^2 & 0 & 0 \\
                          0 & 8xy & 0 & 0 & 12y^2 & 0 \\
                          0 & 0 & 16xy & 0 & 0 & 4x^2+12y^2 
\end{array}\right].\nonumber
\ea   
A diagonalization of ${\bf V}$ yields the eigenvalues
\ba
\lambda_{1,2} &=& 10x^2+14y^2\pm 2\sqrt{x^4+25y^4+54x^2y^2},\nonumber\\
\lambda_{3,4} &=& 14x^2+10y^2\pm 2\sqrt{y^4+25x^4+54x^2y^2},\nonumber\\
\lambda_{5,6} &=& 6(x^2+y^2) \pm 2\sqrt{9x^4-2x^2y^2+9y^4}.\nonumber
\ea
All eigenvalues are non-negative and vanish at the origin $(x=y=0)$.  Thus,
instability may occur only in its vicinity. Though the expansion of a potential
around an invariant manifold is no substitute for the computations of Lyapunov
exponents or monodromy matrices, it is a first step when estimating stability
properties of such manifolds.

\section{Zickendraht's coordinates and invariant manifolds}
\label{zick}
Almost thirty years ago Zickendraht \cite{Zickendraht} introduced a set of
collective coordinates to describe nuclear vibrations and rotations, as well as
their coupling with single particle motion.  We shall discuss to what extent
these coordinates correspond to the ones we introduced in the previous
sections. On one hand this will allow to identify certain vibrational modes of
a many-body system with invariant manifolds. On the other hand we shall also
see that our procedure proposes collective movements that are not of the type
described easily in Zickendraht's coordinates.

Following Zickendraht~\cite{Zickendraht} we write the 
coordinates $\vec{r}_i$ of the
$i^{th}$ particle in the center of mass system as
\be
\label{zickone}
\vec{r}_i=s_{i1}\,\vec{y_1}+s_{i2}\,\vec{y_2}+s_{i3}\,\vec{y_3},\quad i=1,\ldots,N
\ee
where the $\vec{y}_i$ span the inertia ellipsoid and $s_{ik}$ are
non-collective coordinates which for simplicity we shall call 
single-particle coordinates. The newly introduced coordinates $\vec{y}_i$
and $s_{ij}$ are not independent. The constraints are
\ba
\vec{y}_i\cdot\vec{y}_j &=& y_i y_j \delta_{ij},\quad i,j=1,2,3\nonumber\\
\sum_{i=0}^N s_{ij} &=& 0,\quad j=1,2,3\nonumber\\
\sum_{i=0}^N s_{ij}\,s_{ik}&=&\delta_{jk}, \quad j,k=1,2,3.\nonumber
\ea
The first six equations ensure the orthogonality and normalization of the 
principal axis of the
inertia ellipsoid whereas the next three equations fix the origin at the center
of mass system. The last six equations are orthogonality relations of the
single-particle coordinates. In the center of mass system, one may therefore
characterize the $N$-body system by its inertia ellipsoid (e.g. three Euler
angles of the principle axis and three moments of inertia) and $3N-9$ single
particle coordinates. The moments of inertia $I_i$ are related to the
coordinates $y_i$ by
\be
\label{inert}
I_1=m(y_2^2+y_3^2),\qquad I_2=m(y_1^2+y_3^2),\qquad I_3=m(y_1^2+y_2^2),
\ee
where $m$ denotes the mass of the particles. 

It is interesting to determine those configurations, where the motion of
the many-body system may be described in terms of the collective coordinates
$y_i$ only. While such motion would be restricted to {\it some} invariant
manifold in phase space it would not obviously be one of those defined by
eq.~(\ref{mandef}). We may however determine invariant
manifolds~(\ref{mandef}) such that the motion on the manifold changes only the
inertia ellipsoid of the system and hence may be described entirely by
Zickendraht's collective coordinates $y_i$. Two necessary conditions for this a
situation are easily stated. First, the number of coordinates on such invariant
manifold may not exceed six in the general case and three in the case of pure
vibrations. Second, every motion on such an invariant manifold has to change
the inertia ellipsoid of the many-body system. 

For simplicity let us start with the a system of four particles in two
spatial dimensions and the invariant manifold displayed in Fig.~\ref{fig1}, 
i.e.
\ba
\vec{r}_1=\left[\begin{array}{c}x\\y\end{array}\right],\quad
\vec{r}_2=\left[\begin{array}{c}x\\-y\end{array}\right],\quad
\vec{r}_3=\left[\begin{array}{c}-x\\-y\end{array}\right],\quad
\vec{r}_4=\left[\begin{array}{c}-x\\y\end{array}\right],\nonumber
\ea
and the momenta are chosen by replacing $x\to p_x,y\to p_y$.
Computation of the moments of inertia yield the collective Zickendraht
coordinates $y_1=2x$, $y_2=2y$. On the invariant manifold the remaining
coordinates are given by 
$s_{11}=s_{12}=s_{21}=-s_{22}=-s_{31}=-s_{32}=-s_{41}=s_{42}=1/2$. This
shows that every motion on the invariant manifold only changes the moments 
of inertia and therefore decouples from the single-particle motion.  
 
We next consider the example of an eight-body system in three dimensions.
Let 
\ba
\label{conf}
\vec{r}_1=\left[\begin{array}{c}x\\y\\z\end{array}\right],\quad
\vec{r}_2=\left[\begin{array}{c}-y\\x\\z\end{array}\right],\quad
\vec{r}_3=\left[\begin{array}{c}-x\\-y\\z\end{array}\right],\quad
\vec{r}_4=\left[\begin{array}{c}y\\-x\\z\end{array}\right],\quad
\vec{r}_{4+i}=\vec{r}_i(z\leftrightarrow -z)
\ea
denote a configuration restricted to the invariant manifold displayed in
Fig.~\ref{fig2} (a) with $C_{4h}$ symmetry. (The momenta are chosen by
replacing $x\to p_x,y\to p_y,z\to p_z$ in eq.(\ref{conf}).) The moments of
inertia are $I_1=I_2=4m(x^2+y^2)+8mz^2,I_3=8m(x^2+y^2)$ and yield collective
coordinates (\ref{inert}) $y_1^2=y_2^2=4(x^2+y^2),y_3^2=8z^2$. Since the
inertia ellipsoid is symmetric we have a freedom in choosing two of its
principle axis. Using
\ba
\vec{y}_1=2\left[\begin{array}{c} x\\y\\0\end{array}\right],\quad
\vec{y}_2=2\left[\begin{array}{c}-y\\x\\0\end{array}\right],\quad
\vec{y}_3=\sqrt{8}\left[\begin{array}{c}0\\0\\z\end{array}\right].\nonumber
\ea
one obtains constant single-particles coordinates 
$s_{11}=-s_{31}=s_{51}=-s_{71}=s_{22}=-s_{42}=s_{62}=-s_{82}=1/2, 
s_{13}=s_{23}=s_{33}=s_{43}=-s_{53}=-s_{63}=-s_{73}=-s_{83}=1/\sqrt{8}$ 
for the motion on the invariant manifold. Thus, the 
single-particle motion decouples from the collective motion
on the invariant manifold. Similar results hold for the six particle 
configuration displayed in Fig.~\ref{fig2}. 

It is also instructive to consider one counterexample. The configuration
\ba
\vec{r}_1=\left[\begin{array}{c}x\\y\\z\end{array}\right],\quad
\vec{r}_2=\left[\begin{array}{c}-y\\x\\z\end{array}\right],\quad
\vec{r}_3=\left[\begin{array}{c}-x\\-y\\z\end{array}\right],\quad
\vec{r}_4=\left[\begin{array}{c}y\\-x\\z\end{array}\right],\nonumber\\
\vec{r}_5=\left[\begin{array}{c}x\\-y\\-z\end{array}\right],\quad
\vec{r}_6=\left[\begin{array}{c}y\\x\\-z\end{array}\right],\quad
\vec{r}_7=\left[\begin{array}{c}-x\\y\\-z\end{array}\right],\quad
\vec{r}_8=\left[\begin{array}{c}-y\\-x\\-z\end{array}\right],\nonumber
\ea
displays $D_4$ symmetry and
differs from configuration (\ref{conf}) by a shearing motion. Like in the
previous example, the moments of inertia are given by
$I_1=I_2=4m(x^2+y^2)+8mz^2,I_3=8m(x^2+y^2)$ and the ellipsoid of inertia is
symmetric. However, no choice of the principal axis allows to fulfill
eqs.~(\ref{zickone}) with {\it constant} single-particle coordinates $s_{ij}$.
Therefore, single-particle degrees of freedom depend on collective degrees of
freedom and a decoupling does not exist using Zickendraht's coordinate system.
A decoupling is obtained by using the coordinates introduced in
this work. However, the collective motion on the appropriate invariant
manifold does not correspond to pure vibrations or rotations of the
inertia ellipsoid. These findings are interesting e.g. in relation with with 
the magnetic dipole 
mode in nuclei \cite{ARichter} since this type of collective behavior is 
associated with a shearing motion. 

\section{Discussion}
We constructed an orthogonal transformation that maps the Cartesian single
particle coordinates of a many-body system to a new Cartesian coordinate system that
distinguishes collective and single-particle motion. The collective degrees of
freedom govern the motion that is restricted to a low-dimensional invariant
manifold and are decoupled from single-particle degrees of freedom on this
manifold. We have demonstrated that there are several configurations of
few-body systems, where the motion on the invariant manifold corresponds to a
vibration or rotation and may be described in terms of Zickendraht's collective
coordinates, but differs when the collective motion goes beyond that. 
These results are independent of the
details of the Hamiltonian of the $N$-body system, and are entirely determined
by rotational and permutational symmetry.

Using the results of this article as well as those of refs. \cite{PP99,PSW98}
we can draw the following picture: First it is possible that an invariant
manifold is spanned exactly by the vibrational and rotational modes of a
few-body system; second such manifolds may be stable or have small instability
exponents in transversal directions; third the revival probabilities of wave
packets launched on such manifolds are large; last, as a conclusion of these
points we may have a collective motion near the manifold whose damping is
characterized by the decay rate in transversal direction. We also found that
there may be other collective motions; this was displayed in an example of shearing
motion, but there can be others such as breathing modes etc. The coincidence
of Zickendraht coordinates with our collective coordinates depends on particle
numbers; typically they do not span an invariant manifold. This confirms
the well-known fact that in general the collective motion in these coordinates
does not separate rigorously, but only in some adiabatic approximation. 

As we found more general invariant configurations which in turn induce
collective coordinates we may hope that these are useful for approximate
considerations for larger particle numbers that do include the corresponding
invariant manifold in a non-trivial fashion. The construction of appropriate
coordinates is an open problem.

\section*{Acknowledgments}
T.P. acknowledges the warm hospitality at the 
Centro Internacional de Ciencias, Cuernavaca, Mexico, where part of
this work has been done. This work was partially supported by Dept. of
Energy (USA), by DGAPA (UNAM) and by CONACyT (Mexico).

\end{document}